\documentclass[
reprint,
 amsmath,amssymb,
 aps,
]{revtex4-2}

\usepackage{color}
\usepackage[bottom]{footmisc}
\usepackage{graphicx}
\usepackage{dcolumn}
\usepackage{bm}

\begin{document}

\preprint{APS/123-QED}

\title{Doppler effect in TianQin time-delay interferometry}


\author{Lu Zheng}
\author{Shutao Yang}
\author{Xuefeng Zhang}
 \email{zhangxf38@sysu.edu.cn}

\affiliation{MOE Key Laboratory of TianQin Mission, TianQin Research Center for Gravitational Physics $\&$ School of Physics and Astronomy, Frontiers Science Center for TianQin, Gravitational Wave Research Center of CNSA, Sun Yat-sen University (Zhuhai Campus), Zhuhai 519082, China}


\date{\today}

\begin{abstract}
The current design of space-based gravitational wave detectors utilizes heterodyne laser interferometry in inter-satellite science measurements. Frequency variations of the heterodyne beatnotes are predominantly caused by the Doppler effect from relative satellite motion along lines of sight. Generally considered to be outside the measurement band, this Doppler frequency shift appears to have been overlooked in numerical simulations of time-delay interferometry (TDI). However, the potential impact on the {implementation} of TDI should be assessed. The issue is particularly relevant to TianQin that features geocentric orbits, because of strong gravity disturbances from the Earth-Moon system at {frequencies} $<1\times 10^{-4}$ Hz. In this proof-of-principle study, based on high-precision orbital data obtained from detailed gravity field modeling, we incorporate the Doppler shift in the generation of TianQin’s beatnote phase signals. To remove the large-scale Doppler phase drift at {frequencies} $<1\times 10^{-4}$ Hz, we develop a high-performance high-pass filter and consider two possible processing sequences, i.e., applying the filter before or after TDI combinations. Our simulation results favor the former and demonstrate successful removal of the low-frequency gravity disturbances for TianQin without degrading the TDI performance, assuming 10 m pseudo-ranging uncertainty. The filtering scheme can be used in developing the initial noise-reduction pipeline for TianQin. 
\end{abstract}

\maketitle



\section{\label{sec:intro} Introduction}

Gravitational waves (GWs) are excellent information carriers and are expected to provide a unique and new method for observing and studying the universe. With the development of technology, GWs have been detected by the LIGO detectors for the first time in 2015 utilizing laser interferometry \cite{abbott2016observation}. LIGO have an armlength of 4 km, and their most sensitive frequency band is about $10-10^3$ Hz \cite{abbott2009ligo}. Limited by available armlengths and seismic noise, in order to detect GWs in the mHz frequency band, it is necessary to send GW detectors to space. For this reason, space-based GW detector missions such as LISA and TianQin have been proposed \cite{amaro2017laser,Luo2016CQGTianQin}.

The TianQin mission is expected to deploy three identical satellites in a circular orbit with the radius about $10^5$ km of the Earth, which will form an approximately equilateral triangle constellation with an armlength of about $ 1.7\times10^5 $ km \cite{Ye2019OrbitsOptimizing,Tan2020Orientation}. This interference armlength enables the TianQin observatory to have a sensitive detection capability in the 0.1 mHz - 1 Hz frequency band. There are a wide variety of important astrophysical and cosmological sources in this frequency band, including massive binary black holes, extreme mass ratio inspirals (EMRIs), compact galactic binaries and stochastic GW backgrounds \cite{WangHT2019PRDmassiveBH, HuangSJ2020PRDWhiteDwarf, Fan2020PRDEMRI, Mei2021PTEPTQProgress}.

Like other space-based interferometric GW detectors, TianQin suffers from overwhelming laser phase noise due to the coupling with unequal armlengths caused by the relative motion of the satellites. After many theoretical studies and experimental verifications, time delay interferometry (TDI) is considered to be an effective method for deducting laser phase noise \cite{TintoPRD1999TDI, Armstrong1999TDI, Estabrook2000PRDTDI, Tinto2003PRDTDI, Vine2010TDIexperimental, schwarze2019picometer,Tinto2021TDIReviews}. As a data post-processing technique, TDI can construct virtual equal-armlength {interferometers}  to reduce laser phase noise. The LISA team has developed multiple sets of TDI simulation software \cite{rubbo2004forward, Vallisneri2005PRDTDI, petiteau2008lisacode, otto2015PHD, Otto2012CQGTDI, bayle2019simulation}, such as LISA Simulator, Synthetic LISA, LISACode, and LISANode, and the capabilities have been improved over time. As possibly the first simulator, LISA Simulator was based on an old interferometry layout \cite{rubbo2004forward}. Synthetic LISA and LISACode are more in line with the real physical scenario \cite{Vallisneri2005PRDTDI, petiteau2008lisacode}. LISANode is a new {prototype simulator} which includes an up-to-date instrumental configuration \cite{Heinzel2011CQGAuxiliary}, various noise sources, and improved TDI algorithms \cite{bayle2019simulation,bayle2021adapting,Hartwig2021PRDclock}. Important lessons can be learned from LISA. Nevertheless, to handle unique features of TianQin, we have felt the necessities to develop a TDI simulation program specifically for the mission. Some earlier efforts were made, for instances, in studying the science cases of EMRIs \cite{Fan2020PRDEMRI} and in evaluating residual unequal armlengths after TDI \cite{Zhou2021PRDorbital}, and, more recently, in assessing the space plasma noise propagation in TDI \cite{Jing2022plasma}. The new \texttt{MATLAB}-based program is named TQTDI and has been developed in tandem with our orbit simulator TQPOP (TianQin Quadruple Precision Orbit Propagator \cite{Zhang2021PRDEarthMoon}). 

TDI is currently a necessary method for deducting laser phase noise that is several orders of magnitude higher than GW signals in space GW detection, but there are other important effects to be reckoned with in TDI data processing. Because of phase locking offsets and the Doppler shift of the laser frequency due to relative motion between satellites, the beatnote frequencies can vary by $\sim 10$ MHz {(The bandwidth for phase measurement is about 5 to 25 MHz \cite{Barke2015PHD, bayle2023unified})}. In the raw output of phasemeters, small GW-induced phase fluctuations are buried under large phase ramps, but {most of} previous TDI simulations in literature seem to have disregarded the Doppler shift in generating the science beatnote signals. Though the Doppler effect is considered out-of-band especially for LISA, the potential impact on the effectiveness of TDI deserves careful studies. Additionally, the initial noise reduction pipeline should take the issue into account in order to handle realistic data, since it only has access to
total phases or frequencies {\cite{bayle2021adapting,hartwig2021instrumentalPHD}}. 

Proximity of the TianQin satellites to the Earth has its benefits, such as easier data communication, availability of the Global Navigation Satellite Systems, etc. However, the complex gravity environment of the Earth-Moon system also sets a lower bound at $1\times 10^{-4}$ Hz to the detection frequency band of TianQin \cite{Zhang2021PRDEarthMoon, luo2022effect}, which poses a risk of interfering with TDI processing.

In order to explore whether or not the Doppler effect affects TianQin's TDI processing, we need to compare the results of TDI before and after deducting the Doppler shift. Due to the huge magnitude difference between GW signals and the Doppler effect when expressed in phase, double-precision (16 digits) can no longer meet the precision requirement. As a remedy, our simulator has been developed using quadruple-precision arithmetic (34 digits) at necessary places, which is a technique we have previously applied for TQPOP \cite{Zhang2021PRDEarthMoon}. Because the orbital Doppler effect is very close to the detection frequency band of TianQin \cite{Zhang2021PRDEarthMoon,luo2022effect}, the method of deducting it also needs to be carefully designed and tested to ensure that it does not affect TDI performance and GW detection. 

Ground-based GW interferometers already in operation, such as advanced LIGO, also suffer from various environmental noises, among which seismic and Newtonian noises in the raw data is rather high at low frequencies below 10 Hz \cite{aasi2015advanced, fiori2021environmental, nguyen2021environmental}. In order to facilitate subsequent scientific data analysis, one approach adopted by advanced LIGO is using high-pass filter in \texttt{gstlal} calibration pipeline to remove the huge low frequency noise \cite{viets2018reconstructing, cannon2021gstlal}. TianQin's case is somewhat similar but with the complication of TDI. Hence we may also take a similar approach in the pre-data processing for TianQin to remove this orbital noise. 

The paper is organized as follows. In Sec. \ref{sec:Theory} we introduce the cause of the Doppler effect and deduce the TDI formulas with the effect. In Sec. \ref{sec:method}, first we analyze the characteristics of the orbital Doppler effect of the TianQin satellites, which shows the feasibility of using a high-pass filter to deduct the effect. Then we briefly illustrate the design method of the filter used in our research and the simulation tool TQTDI we developed is described. Section \ref{sec:results} shows the simulation results, including the data processing order and the effect of ranging errors, and theoretical deduction of the influence of Doppler effect on TDI processing is also included. Last, conclusions are presented in Sec. \ref{sec:conclusion}.


\section{\label{sec:Theory} Theoretical description}

\begin{figure}[tb]
\includegraphics[width=0.45\textwidth]{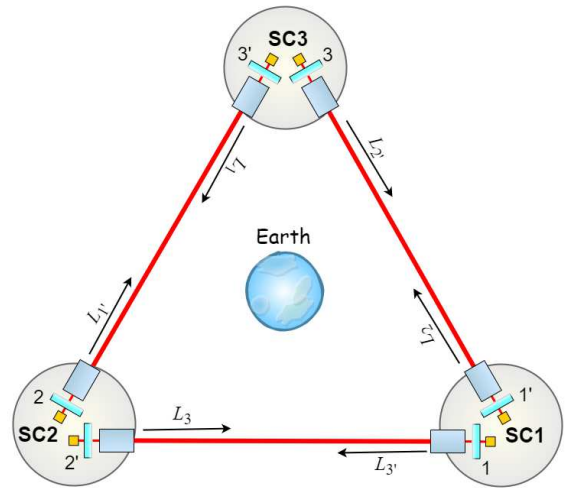}
\caption{\label{fig:Notation} Labeling conventions used in TianQin split-interferometer. }
\end{figure}

The TianQin constellation is shown in Fig. \ref{fig:Notation}, which includes three spacecrafts that are labeled SC$i$ (For $i$ = 1, 2, 3, same below). The lengths of laser links between satellites are denoted as $L_i$ and $L_{i'}$, respectively, according to the counter-clockwise and clockwise propagation directions. Each of the three spacecrafts carries two optical benches (OB) marked as $i$ and $i'$. In the split measurement scheme  {\cite{Otto2012CQGTDI}}, each OB will output three sets of measurement data from the science interferometer, test-mass (TM) interferometer and reference interferometer (The derivation in this paper neglects clock sideband measurements). The measurement data are denoted as $s_i$, $\epsilon_i$, and $\tau_i$, respectively, which contain both GW signals and noises. TDI algorithm is to linearly combine the measurements and their delayed terms to suppress the laser phase noise and retain the GW signals. There are a variety of TDI combinations, such as Michelson combinations X (1st generation), X$_2$ (2nd generation), Sagnac combinations $\alpha$, $\beta$, and so on {\cite{Armstrong1999TDI,shaddock2003data}}. The detailed composition of $s_i$,  $\epsilon_i$, and $\tau_i$ has been deduced in the previous literature \cite{Otto2012CQGTDI}, and for simplicity, here we only include a few crucial physical quantities: GW signals, laser phase noise and Doppler frequency shift in the unit of phase for the derivation of the TDI formula. 

Without loss of generality, we take OB1 as an example to derive the TDI combinations with the Doppler effect. The total phase of a photon originating from the laser source of OB1 at the time $t$ is given as 
\begin{equation}
\Phi_1(t)=\omega_1t+p_1(t),
\end{equation}
where $\omega_1$ is the center laser frequency and $p_1$ is the phase fluctuation of the pre-stabilized laser. The total phase of the laser emitted from SC2 when it reaches SC1 at the time $t$ can be expressed as
\begin{equation}
\Phi_{2}^{\prime}(t)=\omega_{2^{\prime}}^{\prime} t+p_{2'}(t-\tau_{3})+H_{1}(t),
\end{equation}
where $\omega_{2^{\prime}}^{\prime}$ is the center frequency of the laser on OB2’ with the Doppler shift, $p_{2^{\prime}}\left(t-\tau_{3}\right)$ is the phase fluctuation of the laser on OB2’ with the delay along the arm $L_3$, and $H_{1}(t)$ is the phase changes caused by GWs. Hence, the phase difference formed by the interference of the two laser beams on OB1 can be expressed as
\begin{equation}\label{eq:dphi}
\Delta\Phi_1(t)=(\omega_{2'}'-\omega_1)t+p_{2'}(t-\tau_3)-p_1(t)+H_1(t).
\end{equation}
In Eq. (\ref{eq:dphi}), we regard the term $(\omega_{2'}'-\omega_1)t$ as being solely caused by the Doppler effect, and assume that the fixed frequency difference {introduced during phase locking} has been removed. In reality, the beatnote signal measured by the phasemeter contains this Doppler term {($(\omega_{2'}'-\omega_1)t$)}, and it might affect the GW detection due to the complex gravitational field environment of the geocentric orbit. Therefore, we will add this term in the following derivation, and focus on examining it in this work.

By introducing the delay operator $D_if(t)=f(t-\tau_i)$, the science interferometer signals of the two optical benches in SC1 can be written as
\begin{equation}\label{eq:sci}
\begin{split}
s_1&=\phi_{1}^d+D_{3}p_{2'}-p_{1}+H_{1},\\
s_{1'}&=\phi_{1'}^d+D_{2'}p_3-p_{1'}+H_{1'},
\end{split}
\end{equation}
where $\phi_{1}^d=\int(\omega_{2'}'-\omega_1)dt$ and  $\phi_{1'}^d=\int(\omega_{3}'-\omega_{1'})dt$ represent accumulated Doppler phases. The multiplication is converted to an integral here because the frequency of the received laser beam sent by the remote satellite changes due to the relative movement between the satellites. Note that in this short derivation alone we ignore OB displacement noise, clock noise, readout noise, and {other noises here \cite{Otto2012CQGTDI}}, and only consider laser phase noise, the Doppler effect, and GW signals. In this case, the TM and reference interferometer signals become trivial. All other interferometer signals from SC2 and SC3 can be constructed by cyclic permutation of the unprimed and primed indices: 1 → 2, 2 → 3 and 3 → 1 as well as 1’ → 2’, 2’ → 3’ and 3’ → 1’. 

Then, we follow the standard procedure to construct the intermediate TDI variables $\eta_1$ and $\eta_{1'}$ to remove the frequency fluctuations of the primed lasers {\cite{Otto2012CQGTDI, otto2015PHD}}: 
\begin{equation}\label{eq:eta}
\begin{split}
\eta_1&=H_{1}+\phi_{1}^d+D_{3}p_{2}-p_{1},
\\\eta_{1'}&=H_{1'}+\phi_{1'}^d+D_{2}p_{3}-p_{1}.
\end{split}
\end{equation}
Similarly, the other four intermediate variables can be obtained by cyclic permutation of the indices. With these variables, we can construct various TDI combinations to remove the laser noises, and the following is the expression for the Michelson-X combination {\cite{Tinto2021TDIReviews}}:
\begin{equation}\label{eq:TDI_X}
\begin{split}
X=&[D_{2'}D_2-I](\eta_1+D_3\eta_{2'})
\\-&[D_{3}D_{3'}-I](\eta_{1'}+D_{2'}\eta_{3}),
\end{split}
\end{equation}
where $I$ is the identity operator. The above equation, like the derivation in previous literature {\cite{Tinto2021TDIReviews}}, will deduct most of the laser noise, leaving only the residual term:
\begin{equation}\label{eq:dTDI_X}
\delta X^p=D_{3'}D_3D_2D_{2'}p_1-D_2D_{2'}D_{3'}D_3p_1.
\end{equation}
One can see from Eq. (\ref{eq:sci}) to Eq. (\ref{eq:TDI_X}) that the TDI combination works in the same way for both $H_i$ and $\phi_i^d$, and does not normally have an additional suppression on the latter. Therefore, we need to analyze and evaluate TianQin’s Doppler effect jointly with TDI to determine whether or not it affects GW detection, and if so, how to deduct it.


\section{\label{sec:method} Simulation setup}

\begin{figure}[ht]
\includegraphics[width=0.45\textwidth]{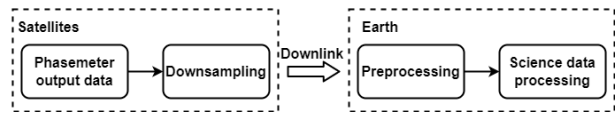}
\caption{\label{fig:data chain} Preliminary TianQin data transmission chain. }
\end{figure}

Figure \ref{fig:data chain} shows a preliminary TianQin data transmission chain. In the satellites, the phasemeter outputs are downsampled to about 10 Hz, and transmitted to the Earth. On the ground, the raw data are preprocessed to reduce the instrument noises through clock synchronization, TDI, and other means {\cite{hartwig2021instrumentalPHD}}. Then the resulting data products are analysed to extract GW signals for science application. Our TQTDI program simulates the steps after the downsampling and before the scientific data processing.

\subsection{Orbital Doppler effect}\label{Doppler}

The Doppler frequency shift is a main component of the phasemeter measurement data. TianQin's floating test masses are subject to complex gravitational perturbations from the Earth, Moon, and other celestial bodies in the solar system. The perturbations are the primary cause of the orbital Doppler effect. In previous work, our team have determined the amplitudes and frequencies of the gravity disturbances for TianQin's orbit, and evaluated its impact on the range acceleration noise \cite{Zhang2021PRDEarthMoon,luo2022effect}. The result shows that the Earth-Moon’s gravity field dominates at low frequencies and is out of the preliminary detection band, which makes it easier to remove the Doppler effect. But the evaluation has also shown the insufficiency of double-precision arithmetic for the numerical simulation because of the enormous dynamic range of the Doppler effect that is about 20 orders of magnitude higher than the noise requirement. To overcome this difficulty, we have developed TDI simulation program supporting quadruple-precision arithmetic in necessary modules, which allows for 34 significant digits in representing simulated data. 

\begin{figure}[t]
\centering 
\begin{minipage}{0.45\textwidth}
\includegraphics[width=\textwidth]{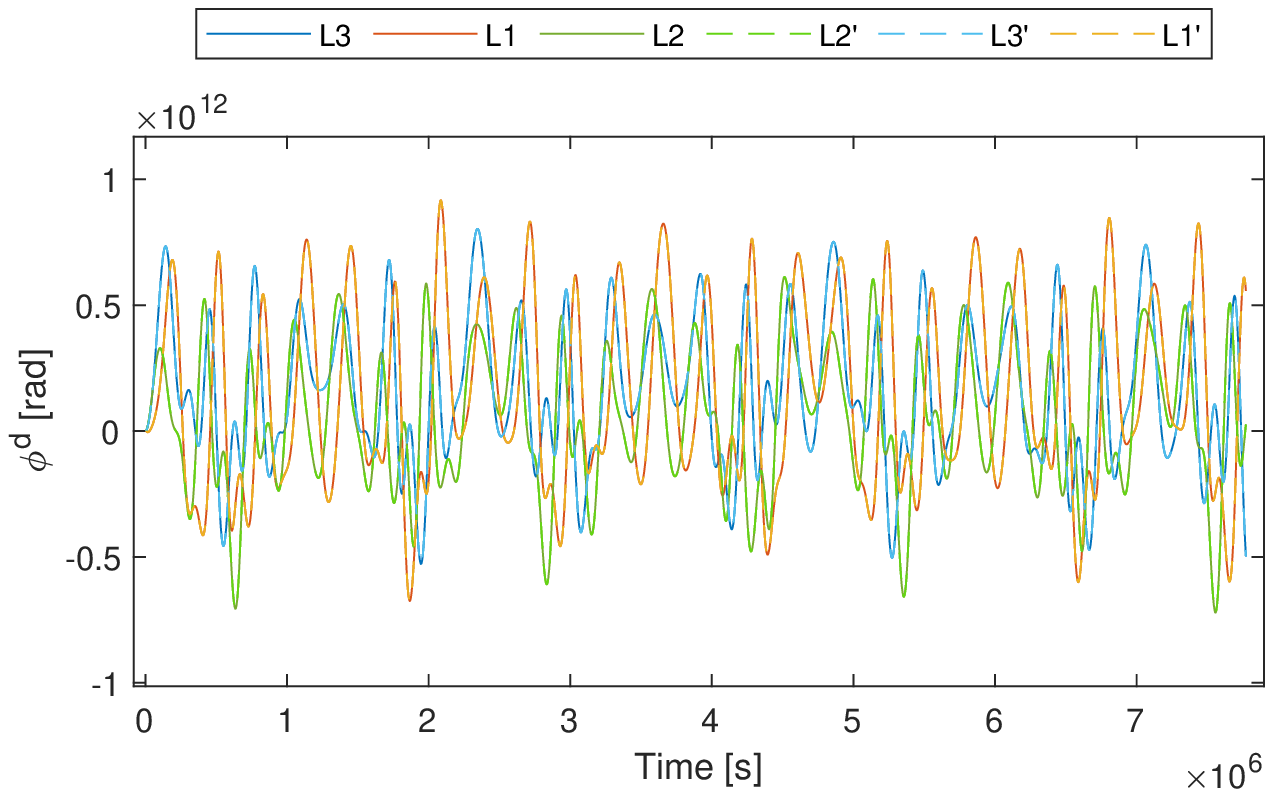}
\end{minipage}
\begin{minipage}{0.45\textwidth}
\includegraphics[width=\textwidth]{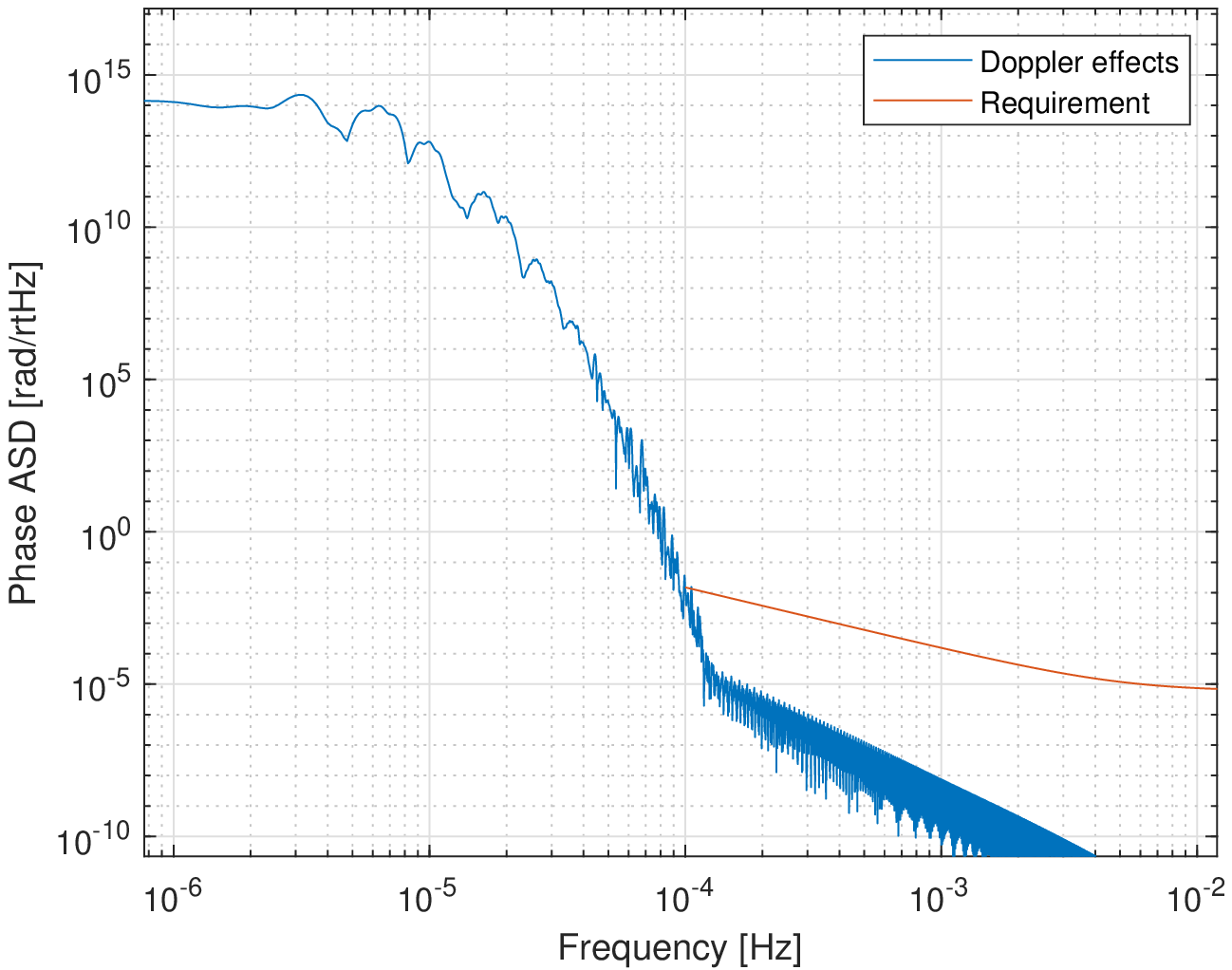}
\end{minipage}
\caption{\label{fig:Doppler} Time and frequency domain diagrams of the orbital Doppler effect of the TianQin satellites (90 days). Above: The phase variations associated with the Doppler effect generated for the six laser links, cf. Fig. \ref{fig:Notation} for notations. Bottom: The amplitude spectral density (ASD) of the Doppler effect and the noise requirement on a single link. }
\end{figure}

TQPOP is a quadruple precision orbit simulation program for TianQin developed by our team \cite{Zhang2021PRDEarthMoon}, which provides a relative truncation error of $<10^{-20}$ and sub-pm/Hz$^{1/2}$ precision.  Various sources of gravitational perturbations are included in the simulator, such as the Earth and Moon’s static gravity, the Earth's solid and ocean tides, the Sun’s point mass and oblateness $J_2$ and so on. Using orbits of the TianQin satellites calculated by TQPOP, the Doppler effect in terms of phase variations can be obtained, and the results are shown in Fig. \ref{fig:Doppler}. One can see that the amplitude spectral density (ASD) is mainly distributed below $1\times 10^{-4}$ Hz. It suggests that a straightforward way to remove the Doppler effect is by high-pass filtering. 

\subsection{High-pass filter design}

A high level of linearity of the phase response \cite{schwarze2019picometer} is needed to meet the stringent requirement for data fidelity in GW science data processing. Thereby a FIR (Finite impulse response) filter that can meet the linear phase requirement is adopted in this work. In order to completely filter out the Doppler effect without affecting GW signals, the FIR filter needs to have the characteristics of large stopband attenuation, narrow transition width, and small peak-to-peak ripple in the passband, which makes the design more difficult. Based on the Parks-McClellan optimal equiripple algorithm \cite{mcclellan1973computer,parks1987digital}, we develop a quadruple-precision high-pass filter program with \texttt{MATLAB}. Applying the filter multiple times in series can remove the Doppler effect by more than ten orders of magnitude while preserving GW signals.

\subsection{Program modules}

\begin{figure}[ht]
\includegraphics[width=0.48\textwidth]{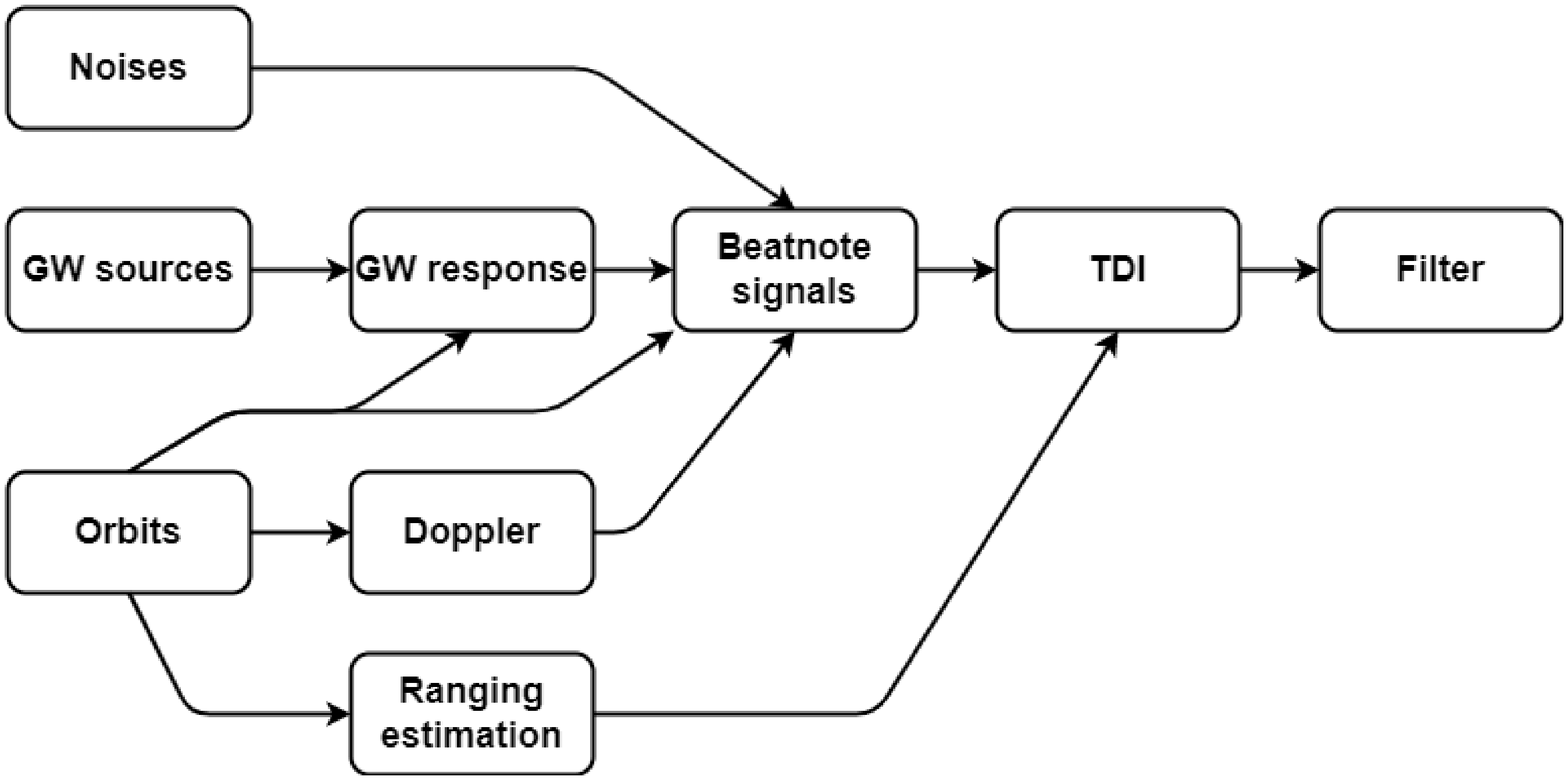}
\caption{\label{fig:TQTDI} Simulation modules of TQTDI. Noises: Generating laser phase noises, TM displacement noises, {readout} noises, and more. GW sources: Providing GW strain based on formulas or file reading, and source information. Orbit: Generating satellite orbit data, calculated by Kepler's formula or importing quadruple-precision orbit data from TQPOP. GW response: Calculating TianQin's responses from satellite orbits and GW sources. Doppler: Calculating the orbital Doppler effect. Ranging estimation: Calculating the laser travel time between two satellites based on the orbits (general relativistic effects ignored), with ranging errors if needed. Beatnote signals: Synthesizing science interferometer signals, TM interferometer signals, and reference interferometer signals of six OBs. TDI: Performing TDI processing on the beatnote signals according to the estimated delay time to cancel the laser phase noise. Filter: Filtering out orbital Doppler effects by a high-pass filter. The order of filtering and TDI may swap. }
\end{figure}

The simulation flow of TQTDI is shown in Fig. \ref{fig:TQTDI}, it can be roughly divided into two parts: raw data generation and preprocessing including TDI and the Doppler effect removal. Two possible preprocessing sequences, i.e., applying the filter before or after TDI combinations, are considered. The raw data, namely the beatnote signals, are dominated by the Doppler effect, but also contains GW signals and various noises. For data generation, one needs to know the satellite orbits and GW sources. The delay time required in the TDI processing can be estimated from the orbits {\cite{otto2015PHD}}.

\section{\label{sec:results} Simulation results}

Using TQTDI, we simulate quadruple-precision raw data with a low sampling rate of 0.02 Hz and a duration of about 90 days. This is chosen partly to save computation time. Then the laser phase noise and Doppler effect are removed by TDI processing and high-pass filtering, respectively. To demonstrate the effectiveness of the preprocessing, a double white dwarf GW signal, with a strain $h = 10^{-18}$ and frequency $f_0$ = 1 mHz, is selected as a test signal in the simulation. Noise parameters are acquired from \cite{Luo2016CQGTianQin}, e.g., 10 Hz/Hz$^{1/2}$ for pre-stabilized lasers. Phase-locking, {which induces a linear growth trend in the output phase data,} has not been added to our simulation, so the detrending is omitted. The results are discussed below. 

\subsection{TDI and filtering without ranging error}

\begin{figure}[t]
\includegraphics[width=0.45\textwidth]{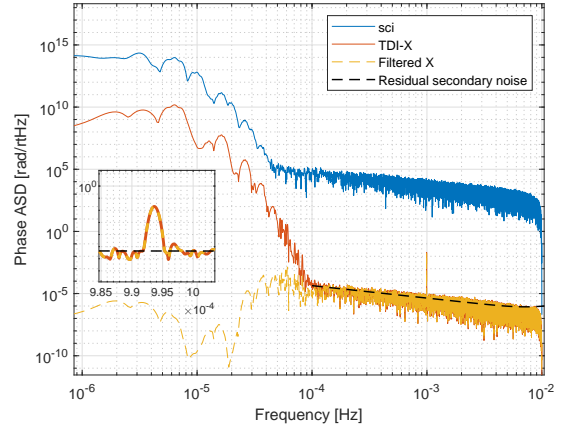}
\caption{\label{fig:TDI_X_f} Phase ASD of the simulation results of 90 days. The blue curve is the science interferometer signal of OB1. The red and yellow curves are TDI-X and TDI-X after high-pass filtering, respectively. The black dashed line is the TDI-X residual secondary noise requirement of TianQin.  {The insert is a zoomed-in plot around $10^{-3}$ Hz and the pick is the test GW signal.} }
\end{figure}

The simulation of this subsection is the ideal situation, that is, one has the perfect information of the absolute ranging of each two satellites. Figure \ref{fig:TDI_X_f} shows the results of TDI processing first and then high-pass filtering. The science interferometer signal (blue curve) shows that the laser phase noise dominates at higher frequencies while the Doppler effect dominates at lower frequencies. Apparently, the double white dwarf GW test signal is completely overwhelmed. After TDI processing, the laser noise is suppressed by about ten orders of magnitude, and the test GW signal appears, as shown by the red curve in the figure. The residual noise at higher frequencies is consistent with the theoretical value of the residual secondary noise after TDI-X \cite{estabrook2000time}, indicating that the TDI-X combination can suppress the laser phase noise below the secondary noise at least in the frequency band of 0.1--1 mHz \cite{Zhou2021PRDorbital}. Note that the expected violation of the noise requirement above 1 mHz \cite{Zhou2021PRDorbital} is not clearly reflected in the 90-day ASD due to non-stationarity of the laser phase noise that is proportional to time-varying residual unequal armlengths \cite{TintoPRD1999TDI}. 

\begin{figure}[ht]
\centering 
\begin{minipage}{0.45\textwidth}
\includegraphics[width=\textwidth]{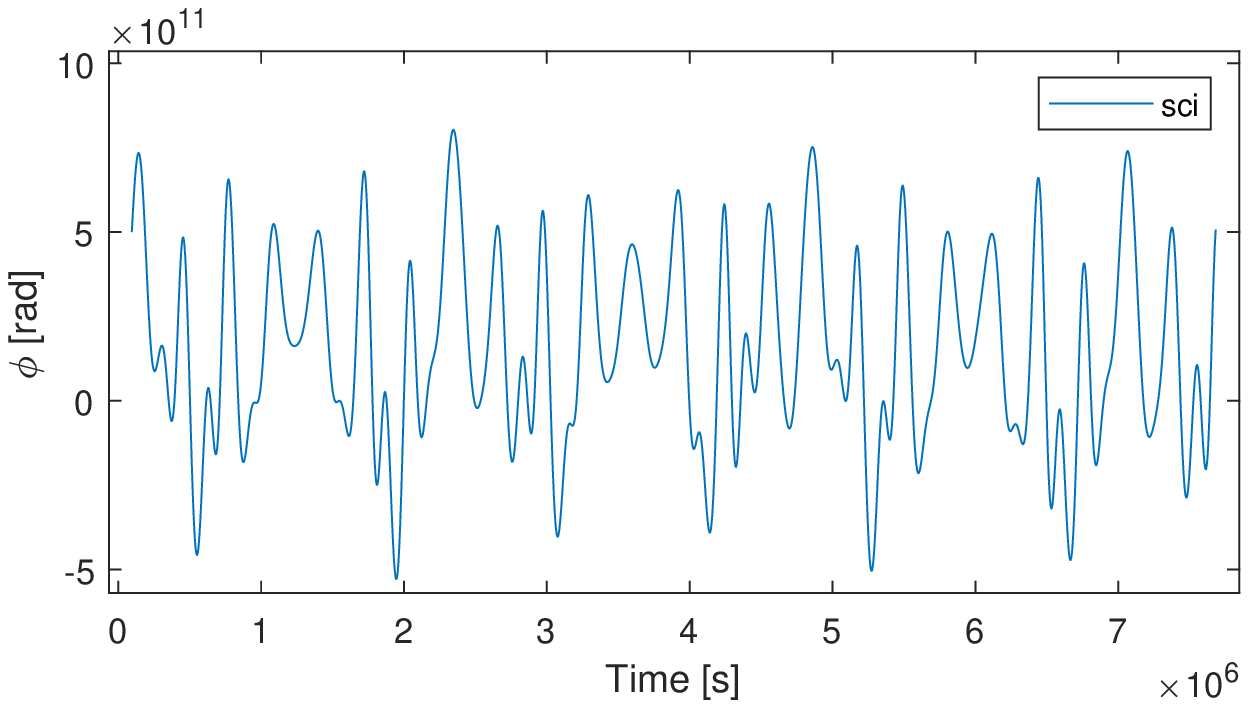}
\end{minipage}
\begin{minipage}{0.45\textwidth}
\includegraphics[width=\textwidth]{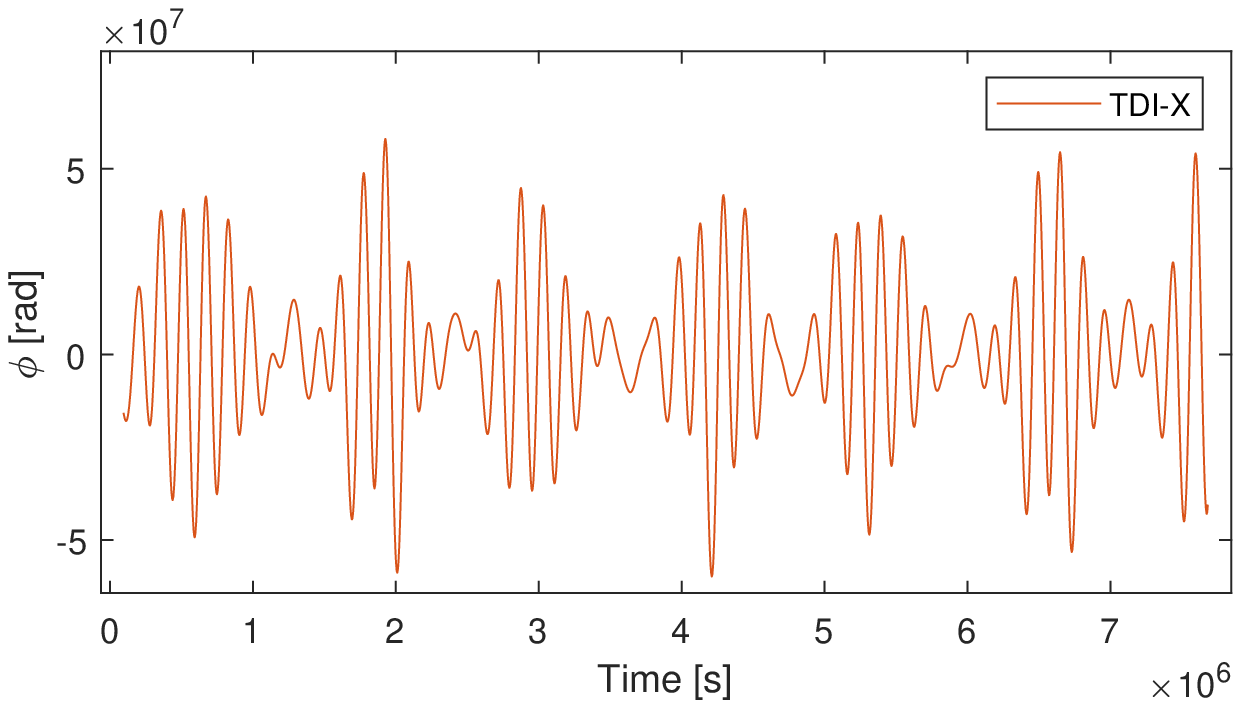}
\end{minipage}
\begin{minipage}{0.45\textwidth}
\includegraphics[width=\textwidth]{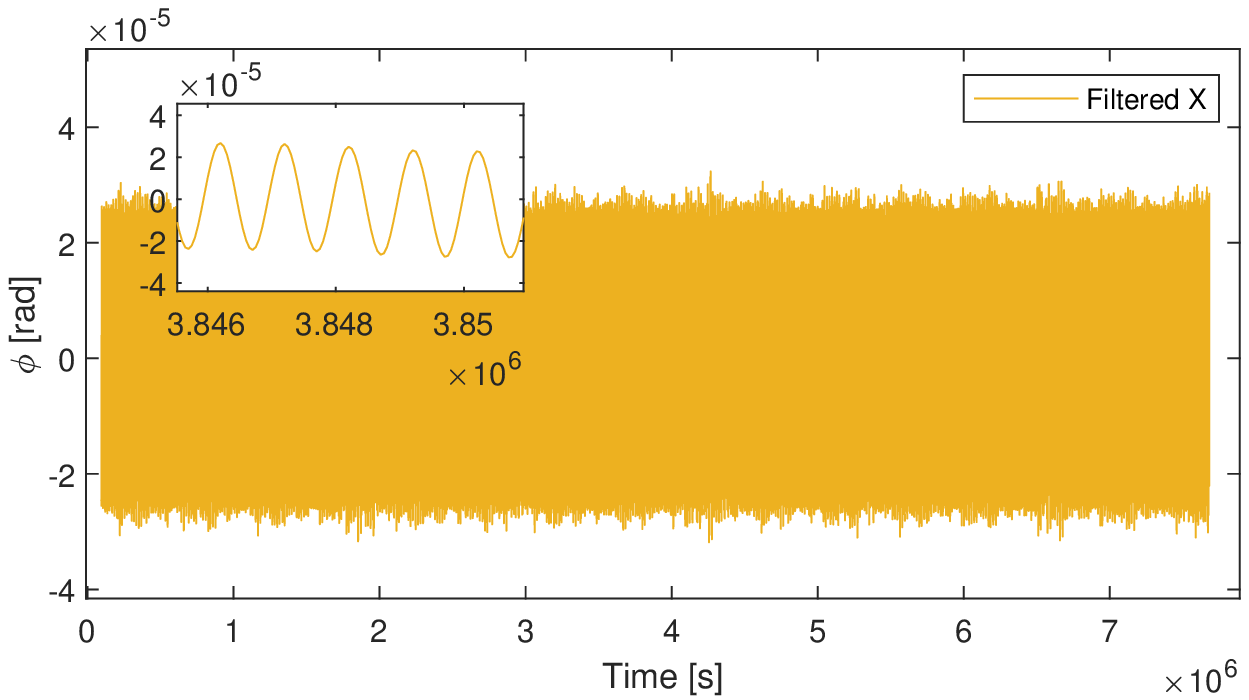}
\end{minipage}
\caption{\label{fig:TDI_X_t} Time domain results of the simulation. The blue, red, and yellow curves show the corresponding time series of the ASD curves in Fig. \ref{fig:TDI_X_f}. The insert of the bottom graph displays an expanded waveform. 
}
\end{figure}

Because of the huge dynamic range of the Doppler effect, quadruple-precision data are used throughout the above simulation, which consume more memory and time than double-precision. In addition, the magnitude of the Doppler effect is much larger than GW signals, which may affect subsequent scientific analysis if not removed. From the first and middle graphs in Fig. \ref{fig:TDI_X_t}, one can only see the characteristics of the orbital motion in the time domain, but no obvious traces of various noises or GW signals. So the output data of TDI processing is high-pass filtered. From the yellow curve in Fig. \ref{fig:TDI_X_f}, it can be seen that the GW signal becomes apparent and the dynamic range of the data is reduced after filtering. Now the data can be converted into double-precision without losing accuracy. Meanwhile, one can also see the sinusoidal GW signal in the time domain, as shown in the bottom graph of Fig. \ref{fig:TDI_X_t}. 

\begin{figure}[ht]
\includegraphics[width=0.45\textwidth]{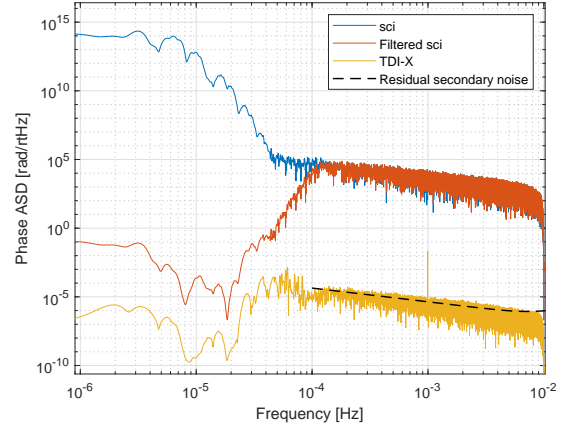}
\caption{\label{fig:TDI_X_f2} 
Phase ASD of the simulation results for filtering before TDI. The blue and red lines are the original science interferometer signal and the filtered signal, respectively. TDI-X from the filtered data is plotted by the yellow line. The residual secondary noise requirement is indicated by black dashed line.}
\end{figure}

The simulation results for exchanging TDI processing and filtering sequence are shown in Fig. \ref{fig:TDI_X_f2}. The laser phase noise is also suppressed below the secondary noise like the previous case in Fig. \ref{fig:TDI_X_f}. Therefore, it can be concluded that when there is perfect knowledge of the armlengths, the sequence of TDI processing and filtering out the Doppler effect does not affect the results.

\begin{figure}[hb]
\includegraphics[width=0.45\textwidth]{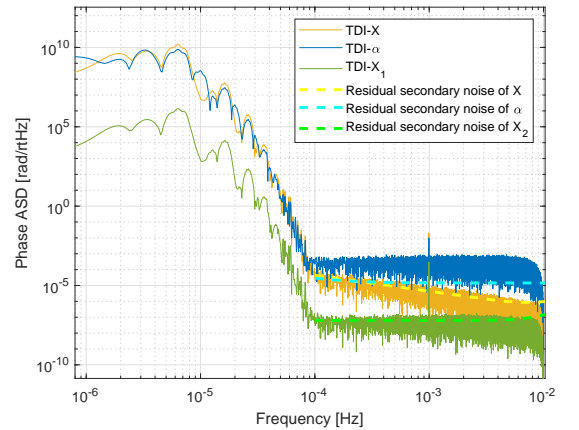}
\caption{\label{fig:TDI_all}  Phase ASDs of different TDI combinations, namely, 1st (yellow) and 2nd (green) generation Michelson combinations and 1st generation Sagnac $\alpha$ combination (blue). The dashed lines represent the corresponding theoretical residual secondary noises. }
\end{figure}

We also simulate several different TDI combinations, and the results of the combinations $\rm X,\  X_2$ and $\alpha $ are shown in Fig. \ref{fig:TDI_all}. By comparing the simulation and theoretical results, one can see that the second generation Michelson combinations can suppress the laser phase noise below the secondary noise in the tested frequency band, but the first generation Sagnac combination cannot. 

\subsection{TDI and filtering with ranging error}\label{sec:TDI_err}

Next, we consider the case with ranging error, and use imperfect knowledge of the armlengths in the simulation. Pseudo-random codes ranging technique \cite{Heinzel2011CQGAuxiliary} is normally used to determine the absolute range between satellites. Here we assume that the ranging errors of the six links are uncorrelated zero-mean Gaussian white noises {\cite{xie2020orbit,hartwig2021instrumentalPHD, bayle2023unified}}, and simulate several scenarios with various noise levels. By adding white noise series with different RMS (Root-Mean-Square) to the laser travel times in TDI, multiple cases with different ranging errors can be simulated. 

\begin{figure}[ht]
\includegraphics[width=0.45\textwidth]{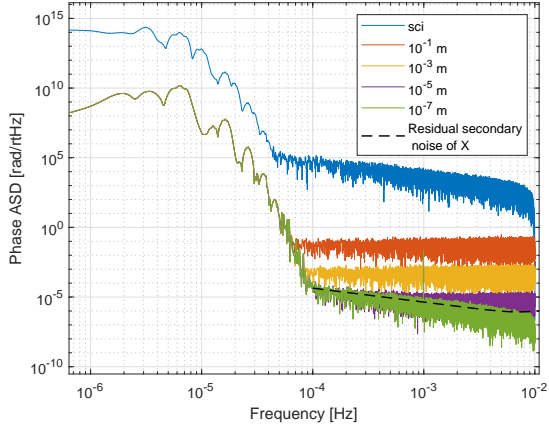}
\caption{\label{fig:TDI_err} The simulation results of TDI processing without removing the Doppler effect in the science interferometer signals, and with ranging errors $ 10^{-1} \ \rm m  , 10^{-3}\  m  , 10^{-5}\  m  , 10^{-7}\  m   $, respectively. The solid blue curve is the science interferometer signal with the Doppler effect, and other colored solid curves are the results of TDI-X with different ranging errors. The residual secondary noise requirement is shown by the black dashed line. 
}
\end{figure}

\begin{figure}[ht]
\includegraphics[width=0.45\textwidth]{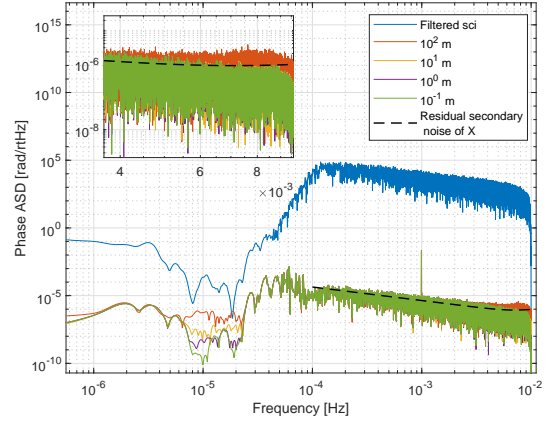}
\caption{\label{fig:TDI_err2} The simulation results of TDI with the Doppler effect filtered out and with the ranging error is $ 10^{2} \ \rm m  , 10^{1}\  m  , 10^{0}\  m  , 10^{-1}\  m$. The high-pass filtered science interferometer signal is plotted by the solid blue curve, and other colored solid curves are the results of TDI-X with different ranging errors. The black dashed line is the residual secondary noise requirement. The insert shows an expanded view.
}
\end{figure}

Figure \ref{fig:TDI_err} shows the simulation results of TDI processing and then filtering, and the ranging error of $ 10^{-1} \ \rm m,\  10^{-3}\  m,\  10^{-5}\  m,\  10^{-7}\  m$ (RMS) are chosen. It can be seen that the green curve coincides with the black dashed line at $>10^{-4}$ Hz, which means the ranging error needs to reach $10^{-7}$ m to make the TDI processing achieve the expected performance. However, in the scenarios without the Doppler effect, namely swapping the order of TDI processing and filtering, the former still shows good suppression even when the ranging error reaches 10 m as displayed in Fig. \ref{fig:TDI_err2}. The discrepancy in the above simulation results implies that an increase in the residual noise may be caused by the Doppler effect not removed in the data. To confirm this, we have the following theoretical derivation. 

When there is delay error,  $ D\phi (t) $ will be $\phi (t-\tau-\delta \tau)$, where $\delta \tau$ represents the time error associated with the ranging error. Then performing a Taylor expansion with respect to $\delta \tau$, one can get the following formula:
\begin{equation}
\phi (t-\tau-\delta \tau) \approx \phi (t-\tau)-\dot\phi (t-\tau)\delta \tau.
\end{equation}
In the scenario where $\dot\phi (t-\tau)$ changes very slowly relative to $\delta \tau$ as in the case of the Doppler effect, we can assume that $ \dot\phi (t-\tau) $ is a constant, denoted by $ \dot\phi$. Applying the Fourier transform, we obtain the error term:
\begin{equation}\label{eq:D_err}
\tilde\phi_{err}={\dot \phi}\,\delta\tilde\tau.
\end{equation}
The above equation shows that the error emerges from the coupling of the derivative of the measured phase $ \dot\phi$ and the time error $\delta \tilde\tau$. The main component of the interferometer signal comes from the Doppler effect before the high-pass filtering, and from laser phase noise after the filtering. In Fig. \ref{fig:TDI_X_f2}, the maximum value of the blue curve is nearly ten orders of magnitude higher than that of the red one. That is to say that $ \dot\phi$ is reduced by nearly ten orders of magnitude after the filtering, which can explain why the ranging error can be relaxed when the Doppler effect is not present.

Using a similar method, one can deduce that for the Doppler effect in TDI-X combination, the error term introduced due to the coupling with the ranging error is
\begin{equation}\label{eq:tdi_err}
\delta X_{err}^d\approx -2\dot\phi_1^d (t)(\delta \tau_2+\delta \tau_{2'})+2\dot\phi_{1'}^d (t)(\delta \tau_3+\delta \tau_{3'}).
\end{equation}
In Eq. (\ref{eq:tdi_err}), we assume the delay time $\tau_i=\tau$ and $ \dot\phi_i^d (t-N\tau)=\dot\phi_i^d (t)$ for $N=1,\ 2,\ 3$. Because $\dot\phi_i^d (t)$ changes quite slowly that $ \dot\phi_i^d (t-N\tau)-\dot\phi_i^d (t)$ is five orders of magnitude lower than $\dot\phi_i^d (t)$. Similarly, considering $\dot\phi_i^d (t)$ as a constant, and assuming that all ranging noises are uncorrelated and have the same frequency characteristics, then in the Fourier domain we obtain
\begin{equation}\label{eq:tdi_err_f}
\delta \tilde X_{err}^d\approx 4\dot\phi^d \delta\tilde \tau.
\end{equation}
For TianQin, $\dot\phi_i^d\sim 10^{7}$ Hz, and by Eq. (\ref{eq:tdi_err_f}), we can estimate that the residual ranging error compatible with the Doppler effect needs to reach about $10^{-7}$ m for TDI-X combination to be below the residual secondary noise over the test frequency band. {However, achieving such precision is currently unfeasible with pseudo-ranging techniques, which can achieve accuracy at the centimeter level \cite{esteban2011experimental, wang2015first, xie2020orbit}.}

Therefore, under the current technology, we recommend removing the Doppler phase drift before performing TDI combinations in the data preprocessing in order to alleviate the requirement on the ranging error. 

\section{\label{sec:conclusion} Conclusion}

In this proof-of-principle study, based on the sub-pm/Hz$^{1/2}$ precision orbits for the TianQin satellites, we develop a quadruple-precision TDI simulation program to account for the large-scale Doppler effect in the science beatnote phase signals. A high-performance high-pass filter compatible with standard TDI processing is designed to remove the Doppler phase drift at low frequencies so that GW signals may emerge in the time domain. Our simulations further demonstrate that TDI processing without removing the Doppler effect (i.e., the orbital noise) beforehand would impose a much higher requirement on the pseudo ranging error, and the explanations are given (see Sec. \ref{sec:TDI_err} and {Appendix}). Hence, we recommend filtering out the Doppler phase drift before performing TDI combinations in TianQin's initial noise-reduction pipeline under the existing pseudo ranging capability. With this scheme, the Earth-Moon's gravity disturbance at {frequencies} $<1\times 10^{-4}$ Hz is expected to have little impact on the GW detection at {frequencies} $>1\times 10^{-4}$ Hz. 

As an added note, the study has assumed the two-way pseudo-ranging for implementing TDI as a baseline for TianQin. However, this is not the only option, and may be complemented by, e.g., TDI ranging \cite{tinto2005time, francis2015tone, page2021bayesian}. Thereby it would be interesting to consider this latter scenario in future work. Moreover, one might also consider finding possible usages for the low-frequency gravity-field information discarded by the filtering.


\begin{acknowledgments}
The authors thank Yi-De Jing, Chengjian Luo, Bobing Ye, Jun Luo and anonymous referee for helpful discussions and comments. X. Z. is supported by the National Key R\&D Program of China (Grant No. 2020YFC2201202 and 2022YFC2204600). 
\end{acknowledgments}


\appendix
{
\section{Derivation of Eqs. (\ref{eq:tdi_err}) and (\ref{eq:tdi_err_f}) }
The result of the Doppler term after being processed by TDI-X can be obtained by replacing the $\eta_i$ in Eq. (\ref{eq:TDI_X}) with $\phi^d_i$. When there are ranging errors, it can be written as
\begin{equation}
	\begin{split}
		X^d&=\phi_1^d(t-\tau_2-\tau_{2'}-\delta\tau _2-\delta\tau_{2'})-\phi_{1}^d(t)
		\\&+\phi_{2'}^d(t-\tau_3-\tau_2-\tau_{2'}-\delta\tau _3-\delta\tau _2-\delta\tau_{2'})
		\\&-\phi_{2'}^d(t-\tau_3-\delta\tau _3)
		\\&-\phi_{1'}^d(t-\tau_{3'}-\tau_3-\delta\tau _3-\delta\tau_{3'})+\phi_{1'}^d(t)
		\\&-\phi_3^d(t-\tau_{2'}-\tau_{3'}-\tau_3-\delta\tau_{2'}-\delta\tau _3-\delta\tau_{3'})
		\\&+\phi_3^d(t-\phi_{2'}-\delta\tau_{2'})
	\end{split}
\end{equation}
Using Taylor expansion, we get
\begin{equation}\label{key}
	\begin{split}
		X^d&\sim \phi_1^d(t-\tau_2-\tau_{2'})-\dot{\phi_1^d}(t-\tau_2-\tau_{2'})(\delta\tau _2+\delta\tau_{2'})
		\\&-\phi_{1}^d(t)
		+\phi_{2'}^d(t-\tau_3-\tau_2-\tau_{2'})
		\\&-\dot{\phi_{2'}^d}(t-\tau_3-\tau_2-\tau_{2'})(\delta\tau _3+\delta\tau _2+\delta\tau_{2'})
		\\&-\phi_{2'}^d(t-\tau_3)+\dot{\phi_{2'}^d}(t-\tau_3)(\delta\tau _3)
		\\&-\phi_{1'}^d(t-\tau_{3'}-\tau_3)+\dot{\phi_{1'}^d}(t-\tau_{3'}-\tau_3)(\delta\tau _3+\delta\tau_{3'})
		\\&+\phi_{1'}^d(t)
		-\phi_3^d(t-\tau_{2'}-\tau_{3'}-\tau_3)
		\\&+\dot{\phi_3^d}(t-\tau_{2'}-\tau_{3'}-\tau_3)(\delta\tau_{2'}+\delta\tau _3+\delta\tau_{3'})
		\\&+\phi_3^d(t-\tau_{2'})-\dot{\phi_3^d}(t-\tau_{2'})(\delta\tau_{2'})
	\end{split}
\end{equation}
Then the error term introduced by the coupling of Doppler effect and ranging error is:
\begin{equation}\label{eq:tdi_err1}
	\begin{split}
		\delta X^d_{err}&= -\dot{\phi_1^d}(t-\tau_2-\tau_{2'})(\delta\tau _2+\delta\tau_{2'})
		\\&-\dot{\phi_{2'}^d}(t-\tau_3-\tau_2-\tau_{2'})(\delta\tau _3+\delta\tau _2+\delta\tau_{2'})
		\\&+\dot{\phi_{2'}^d}(t-\tau_3)(\delta\tau _3)
		+\dot{\phi_{1'}^d}(t-\tau_{3'}-\tau_3)(\delta\tau _3+\delta\tau_{3'})
		\\&+\dot{\phi_3^d}(t-\tau_{2'}-\tau_{3'}-\tau_3)(\delta\tau_{2'}+\delta\tau _3+\delta\tau_{3'})
		\\&-\dot{\phi_3^d}(t-\tau_{2'})(\delta\tau_{2'})
	\end{split}
\end{equation}
First we assume the delay time $\tau_i=\tau$ and $ \dot\phi_i^d (t-N\tau)=\dot\phi_i^d (t)$ for $N=1,\ 2,\ 3$. Furthermore, because the Doppler effect on the same arm is approximately equal, we have
\begin{equation}
	\begin{split}
		\dot\phi_1^d (t)\approx\dot\phi_{2'}^d (t)
		\\ \dot\phi_{1'}^d (t)\approx\dot\phi_{3}^d (t)
	\end{split}
\end{equation}
In this way, Eq. (\ref{eq:tdi_err1}) can be approximated as Eq. (\ref{eq:tdi_err}). By assuming $\delta\tilde \tau_2=\delta\tilde \tau_{2'}=\delta\tilde \tau_3=\delta\tilde \tau_{3'}=\delta\tilde \tau $, the Fourier transform of Eq. (\ref{eq:tdi_err}) yields Eq. (\ref{eq:tdi_err_f}).
}
\\

\section{Verification of Eqs. (\ref{eq:tdi_err}) and (\ref{eq:tdi_err_f}) }\label{sec:appe}

\begin{figure}[ht]
\centering 
\begin{minipage}{0.45\textwidth}
\includegraphics[width=\textwidth]{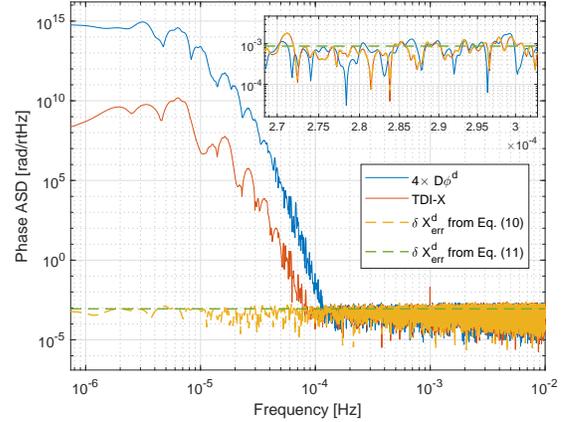}
\end{minipage}
\caption{\label{fig:err} Comparison between theoretical calculation from Eqs. (\ref{eq:tdi_err}, \ref{eq:tdi_err_f}) and numerical simulation. The blue and red solid lines simulated by TQTDI are the ASD of 4 times the Doppler term $\phi_1^d(t)$ with the delay $\tau_2$ and TDI-X combination, both with a ranging error of $10^{-3}$ m, respectively. Yellow dashed line is the ASD calculated from Eq. (\ref{eq:tdi_err}). Taking the RMS of the derivative of each Doppler term as the value of $\dot \phi ^d$, and substituting it into Eq. (\ref{eq:tdi_err_f}) give the green dashed line. }
\end{figure}

Here we verify Eqs. (\ref{eq:tdi_err}) and (\ref{eq:tdi_err_f}) with numerical simulations. The two solid lines in Fig. \ref{fig:err} are the simulation results, and the two dashed lines are the results calculated by Eq. (\ref{eq:tdi_err}) and Eq. (\ref{eq:tdi_err_f}), respectively. As can be seen from  Eq. (\ref{eq:D_err}) and Eq. (\ref{eq:tdi_err_f}), the error introduced by TDI-X is 4 times that of a single delay operator. So we simulate $ 4D_2\phi^d_1(t) $ and TDI-X combination with a ranging error of $10^{-3}\ \rm m $, of which the ASDs are shown by the blue and red solid lines in Fig. \ref{fig:err}. It can be seen that their amplitudes are the same at $>1\times 10^{-4}$ Hz, and are consistent with the theoretical calculation of the dashed lines. The zoomed-in image below shows that the red solid line and the yellow dashed line coincide, and therefore it is shown that the approximations in the deriving Eqs. (\ref{eq:tdi_err}) and (\ref{eq:tdi_err_f}) are reasonable. 

\bibliography{refs}

\end{document}